\documentclass[12pt]{elsart}
\usepackage{amsfonts,amsmath}
\usepackage{times,graphicx}
\usepackage{feynmf}

\newcommand{\bs}[1]{\boldsymbol{#1}}
\newcommand{\vc}[1]{\mathbf{#1}}
\newcommand{\tr}{\mathop{\rm Tr}\nolimits}

\begin{document}

\begin{frontmatter}
\title{ Supersymmetry approach in the field
theory of ergodicity breaking transitions\thanksref{intas} }
\thanks[intas]{Partially supported 
by INTAS under grant 97--1315}
\thanks[email]{e--mail: kisel@elit.chernigov.ua}
\author{Alexei D. Kiselev\thanksref{email}}
\address{
Department of Pure {\slshape \&} Applied Mathematics,\\ 
Chernigov State Technological University,
Shevchenko Street 95, 250027 Chernigov UKRAINE}

\date{}

\begin{abstract}
The supersymmetry (SUSY) self--consistent 
approximation for the model of non--equi\-lib\-rium thermodynamic 
system with quenched disorder 
is derived from the dynamical action 
calculated by means of 
generalized second Legendre transformation technique.
The equations for
adiabatic and isothermal susceptibilities, memory 
and field induced parameters 
are obtained on the basis of asymptotic analysis of
dynamical Dyson equations.
It is shown that the marginal stability condition
that defines the critical point is governed by
fluctuations violating fluctuation--dissipation theorem (FDT). 
The temperature of ergodicity breaking transition
is calculated as a function of
quenched disorder intensities.
Transformation of superfields related to the mapping between
an instanton process and the corresponding causal solution
is discussed.

\end{abstract}


\end{frontmatter}

\section{Introduction}

In recent years significant study has been given to
the microscopic theory of non--equilibrium thermodynamic systems with
quenched disorder that reveal non--ergodic behavior 
and exhibit memory effects.
Spin glasses~\cite{Parisi1,Hertz} and random
heteropolymers~\cite{Sfatos}, 
that received the most of attention, provide the well known examples of such
systems. Procedure of the averaging over disorder is at the heart of
theoretical approaches developed for the description of the systems. 

Thus the bulk of static theories are based on the replica method
firstly introduced in~\cite{Anderson} or employ the methods of random
matrix theory (RMT) (see~\cite{Mehta} for review). In addition, 
the powerful
supersymmetry approach by Efetov~\cite{Efetov}, where generating
functional describing statistics of the density of states is mapped on
the nonlinear supermatrix $\sigma$--model, has had RMT as a key
ingredient of its development.

The starting point of dynamical theories is stochastic dynamics
defined in terms of the corresponding stochastic equations. A wide
range of problems can be formulated in such a way: kinetics of
ordering~\cite{Bray}, non--equilibrium dynamics of spin--glass
systems~\cite{Hertz,Horner,Kurchan1} and so on.
When the stochastic dynamics is governed by Langevin equations, the
generating functional of the stochastic problem can be written as a
functional integral~\cite{MRS,Dominicis} (MRS formalism) thus allowing
for the averaging over disorder at the very beginning of calculations.
It has led to the field theoretic formulation of the problem, so one
could take advantage of using the machinery of the field theory.
The theory is known can represented in the supersymmetric form that
reveals hidden supersymmetry (SUSY) of the stochastic
problem~\cite{Tsvelik,Parisi2}.

In this paper SUSY formalism in the superfield representation is
employed to study ergodicity breaking transition in the model
thermodynamic system with quenched disorder. It implies that 
generating
functional of Langevin dynamical system is represented as functional
integral over superfields with Euclidean action by means of
introducing Grassmann anticommuting variables. These variables and
their products serve as a basis for superfields which components
include Grassmann fields in addition to  the usual real--valued fields. 
The correlation functions of the superfields (supercorrelators) then 
encode physically
relevant information on observables in components of the correlators 
that are the autocorrelator of the order parameter field $C$  and
the response functions $G_{\pm}$ 
(advanced and retarded Green functions).
The analysis
is based on a system of dynamical Dyson equations
supplemented with the equations for averaged order parameter
provided the distribution of quenched random variables is not symmetric.
It was pointed out in~\cite{Kurchan2} that, 
in the mode--coupling approximation,
the Dyson equations can be regarded as
Euler--Lagrange equations for the functional of 
supercorrelators known as dynamical action and
this functional bears striking similarity with the replica expression
for the free energy.

In our study the dynamical action is shown to be 
a second Legendre transformation~\cite{Vasil'ev} of 
the free energy.
So, it depends on both averaged order parameter
and two--time correlation functions and can be calculated
on the basis of suitably defined diagrammatics.
On the technical side, since algebraic structure of the superfield
representation is directly related to the underlying symmetry, we have
the reduction of the number of diagrams to be taken into consideration
in the perturbation theory.
From the other hand, SUSY formalism allows ergodicity breaking transition
be interpreted as a dynamical symmetry breaking transition.
Indeed, in the SUSY language, the well--known "causality" condition
and fluctuation--dissipation theorem (FDT) immediately follow 
from the corresponding Ward identities~\cite{Kurchan1,Justin}.
Below the critical temperature FDT and time translational invariance
are dynamically violated that result in the appearance of
anomalous solutions to the Dyson equations.
In particular, the latter include the case, where the system being 
in a non--ergodic state is
characterized by a very slow relaxation. This phenomenon is known as
aging and,  
as it has been demonstrated for a number of solvable
models~\cite{Cugliandolo1,Franz,Cugliandolo2}, 
the effect is a consequence of trapping in metastable
attractors. 

In this article, leaving aside detailed study of aging,
we shall perform asymptotic and stability
analysis of dynamical Dyson equations to
characterize the transition  
in terms of asymptotic quantities such as 
adiabatic and isothermal susceptibilities, memory and
field induced parameters. This approach is applied to
the simple model of thermodynamic system with quenched
external field and two body interaction.
Note that the
second Legendre transformation technique plays the unifying role
in this theory, so that the SUSY based theory
can
be directly employed for the study of ergodicity breaking transitions in 
different disordered systems such as random heteropolymers~\cite{Sfatos} and
filled nematic liquid crystals~\cite{Kreuzer}. 

Layout of the paper is as follows.

In Sec.~2 the formalism of 
SUSY approach is briefly outlined.
Second Legendre transformation for disordered systems
is introduced in Sec.~3.
In Sec.~4 
the model of non--equilibrium thermodynamic system with quenched disorder is
studied on the basis of 
asymptotic analysis of the Dyson equations.
In high temperature region
it is found that the relevant parameters are the static 
susceptibility $\chi$ and
the field induced parameter $q_h$. The latter is due to the presence
of random field that affects asymptotics of the autocorrelator $C(t)$.
Equations for $\chi$ and $q_h$ combined with the marginal stability condition
define the temperature of ergodicity breaking $T_{c}$.
Dependencies of $T_c$ on the quenched disorder
intensities are calculated.
It is shown that 
the low temperature region $T<T_c$ can be described in terms of
adiabatic $\chi_a$ and isothermal $\chi$ susceptibilities,
the dynamical Edwards--Anderson memory parameter $q$
and the field induced parameter $q_h$, so that
the role of
an order (non--ergodicity) parameter plays 
the difference $\Delta q=q-q_h$.
Discussion of numerical results and concluding remarks are given in
Sec.~5. Appendix~A details the remark that the superfield representation
induced by the shift in time, $t\to t-\bar\theta\theta$, leads to the
mapping between an instanton process and normal downhill motion.

\section{General SUSY formalism}

In this section we sketch the general formalism of SUSY
based theory of a non--equilibrium thermodynamic system.
It serves as an introductionary part of the paper and gives
some details on the results used in subsequent sections.
For definiteness, in what follows we use the lattice designations,
so that the field $\eta_{i}(t)$ defined on sites of the lattice 
(the sites are labelled with the index $i$) gives configuration
of order parameter at the instant of time $t$. 

Relaxational dynamics of the order--parameter field
in the presence of thermal noise is governed
by the Langevin equation: 
\begin{equation}
\dot{\eta}_{i}= - \frac{\delta V}{\delta\eta_{i}} +
\zeta_{i}(t),
\label{eq:1}
\end{equation}
where
the relaxation constant is absorbed by
suitable rescaling of time and $\zeta$. 
The thermodynamic potential $V$ is assumed to be a $t$--local
functional
\begin{equation}
V\{\eta\}=\int \d t\,
V(\eta),
\label{eq:2}
\end{equation}
and $\zeta_{i}(t)$ are Gaussian
stochastic functions subjected to the white noise
conditions:
\begin{equation}
\langle \zeta_{i}(t)\zeta_{j}(0)\rangle_{\zeta}=
2T \delta_{ij} \delta(t),\qquad
\langle\zeta_{i}\rangle_{\zeta}=0,
\label{eq:3}
\end{equation}
where $T$ is the temperature.

\subsection{Generating functional in superfield representation}

The starting point of the MRS formalism~\cite{MRS,Dominicis} is 
the generating functional for correlation functions
of the stochastic problem written in the form:
\begin{equation}
  \label{eq:3a}
Z\{u\}=\left\langle
\int \prod_{i} D\eta_{i}\det\left(\frac{\delta L(\eta)}{\delta\eta}\right)
\delta(L(\eta))
\exp\left(\int\text{d}t u_{i}\eta_{i}\right)
\right\rangle_{\zeta}\, ,  
\end{equation}
where 
$\delta(\cdot)$ is the delta function,
$L(\eta)=-\partial_t\eta-\delta V/\delta\eta +\zeta$
and $\langle\cdot\rangle_{\zeta}$ denotes averaging over noise
realizations.
(Hereafter the functional notations will be adopted 
assuming summation over repeated indexes and integration
over repeated non-discrete arguments.)

Exponentiating the delta function through 'Lagrange multipliers'
$\varphi_{i}(t)$ and the Jacobian functional
determinant through Grassmann fields (ghosts)
$\psi_{i}(t),\,\bar{\psi}_{i}(t)$ and averaging away the noise 
we obtain functional integral representation
for the generating functional (see~\cite{Hochberg} for recent discussion of
the corresponding steps)
\begin{equation}
Z\{u\}=\int \prod_{i} D\eta_{i}D\varphi_{i}
D\psi_{i}D\bar{\psi}_{i}
\exp\{-S+\int\d t u_{i}\eta_{i} \}
\label{eq:4}
\end{equation}
where the action reads
\begin{equation}
S=\int \d t { L},
\label{eq:5}
\end{equation}
\begin{equation}
{ L}=-T\varphi_{i}^{2}+
\varphi_{i}\left\{
\dot{\eta}_{i}+
\frac{\delta V}{\delta\eta_{i}}
\right\}
-\bar{\psi}_{i}\left[
\delta_{ij}\frac{\partial}{\partial t}+
\frac{\delta^{2} V}{\delta\eta_{i}\delta\eta_{j}}
\right]\psi_{j}
\label{eq:6}
\end{equation}

Introducing the superfields
\begin{equation}
\phi_{i}(z)\equiv\phi_{i}=
\eta_{i}+\bar{\theta}\psi_{i}+
\bar{\psi}_{i}\theta + \bar{\theta}\theta\varphi_{i},
\qquad z\equiv\{t,\bar{\theta},\theta\}\equiv\{t,\bs{\theta}\},
\label{eq:7}
\end{equation}
where $\bar{\theta}$ and $\theta$ are anticommuting Grassmann variables,
and substituting
$\{\bar{\theta},\theta\}\to
\{T^{-1/2}\bar{\theta},T^{-1/2}\theta\}$ 
(or, alternatively,
$\{\bar{\psi}_{i},\psi_{i}\}\to
\{T^{1/2}\bar{\psi}_{i},T^{1/2}\psi_{i}\}$,
$\varphi_{i}\to T\varphi_{i}$ 
)
we derive the action as a
functional of superfields
\begin{equation}
S=\frac{1}{T}\int \d z\, { L}=\frac{1}{T}
\int\d t\d^{2}\bs{\theta}\, { L},
\label{eq:8}
\end{equation}
\begin{equation}
{ L}=\bar{D}\phi_{i}D\phi_{i}+V(\phi),
\label{eq:9}
\end{equation}
where $\d^{2}\bs{\theta}\equiv\d\theta\d\bar{\theta}$
and
\begin{equation}
\bar{D}=\frac{\partial}{\partial\theta},\,\,
D=\frac{\partial}{\partial\bar{\theta}}-\theta
\frac{\partial}{\partial t}.
\label{eq:10}
\end{equation}

It is not difficult to show that the operators $\bar{D}$ and $D$ enjoy
the following properties:
\begin{equation}
D^{2}=\bar{D}^{2}=0,\,\,
\{D,\bar{D}\}=-\frac{\partial}{\partial t},\,\,
[D,\bar{D}]^{2}=\frac{\partial^{2}}{\partial t^2},
\label{eq:11}
\end{equation}
where the curly brackets stand for anticommutator,
and the kinetic term in the action
$\displaystyle\int\d z\bar{D}\phi_{i}D\phi_{i}$
can be written in the form
$$
\frac{1}{2}\int\d z \phi_{i}D^{(2)}\phi_{i}\, ,\quad
D^{(2)}\equiv[D,\bar{D}]=
-2\frac{\partial^{2}}{\partial\theta\partial\bar{\theta}}+
\left(1
-2\theta\frac{\partial}{\partial\theta}
\right)
\frac{\partial}{\partial t}\, .
$$
Note that the form of superfield representation (\ref{eq:7})
is fixed up to the change of the basis in the space of superfields, so
that the corresponding transformation would give another
representation of the SUSY group.
Interestingly,
as it is shown in Appendix~A, alternative 
superfield representation, generated by the transformation:
$\phi(t,\bs{\theta})\to\phi(t-\bar\theta\theta,\bs\theta)$,
can be used to construct the mapping between an instanton
process and the corresponding causal solution inverted in time.

\subsection{Symmetries and Ward identities for 2--point functions}

The action by Eqs.~(\ref{eq:8},\ref{eq:9}) is invariant under the action
of the group of supersymmetry~\cite{Kurchan1,Kurchan2}
with generators given by
\begin{equation}
\bar{D}^{\prime}=\frac{\partial}{\partial\bar{\theta}},\qquad
D^{\prime}=\frac{\partial}{\partial\theta}+\bar{\theta}
\frac{\partial}{\partial t},\qquad
\frac{\partial}{\partial t}.
\label{eq:12}
\end{equation}
As a consequence, the correlation function
$\mathbf{G}(z_{1},z_{2})=\langle\phi(z_1)\phi(z_2)\rangle$ (for brevity, the
indexes of superfields are suppressed) meets a set of 
Ward identities provided that the supersymmetry is not 
broken~\cite{Kurchan1}. 
Clearly, the invariance under translations in time implies that
$\mathbf{G}(z_1,z_2)$ depends only on $t=t_1-t_2$. Another two identities are
of special interest to us:
\begin{equation}
(\bar{D}_{1}^{\prime}+\bar{D}_{2}^{\prime})
\mathbf{G}(z_1,z_2)=0,
\label{eq:13}
\end{equation}
\begin{equation}
(D_{1}^{\prime}+D_{2}^{\prime})
\mathbf{G}(z_1,z_2)=0,
\label{eq:14}
\end{equation}
where
 $$
\bar{D}_{i}^{\prime}=\frac{\partial}{\partial\bar{\theta}_{i}},\,
D_{i}^{\prime}=\frac{\partial}{\partial\theta_{i}}+\bar{\theta}_{i}
\frac{\partial}{\partial t_{i}}.
$$
Eq.~(\ref{eq:13}), known as "causality condition", 
implies that the correlator
is of the following form
\begin{equation}
\vc{G}(z_1,z_2)=C(t_1,t_2)+
\left(\bar{\theta}_{1}-\bar{\theta}_{2}\right)
\left(G_{+}(t_1,t_2)\theta_1-
G_{-}(t_1,t_2)\theta_2
\right)\, ,
\label{eq:15}
\end{equation}
where
\begin{subequations}
\begin{eqnarray}
C(t_1,t_2)&=&\langle\eta(t_1)\eta(t_2)\rangle\, ,
\label{eq:16a}\\
G_{+}(t_1,t_2)&=&\langle\varphi(t_1)\eta(t_2)\rangle=
\langle\bar{\psi}(t_1)\psi(t_2)\rangle\, ,
\label{eq:16b}\\
G_{-}(t_1,t_2)&=&\langle\eta(t_1)\varphi(t_2)\rangle=
-\langle\psi(t_1)\bar{\psi}(t_2)\rangle\, .
\label{eq:16c}
\end{eqnarray}
\end{subequations}
Thus the identity (\ref{eq:13}) allows the correlators of
Grassmann fields $\bar{\psi}$ and $\psi$ be expressed in terms of
response functions.

Algebraic structure of the correlator can be conveniently emphasized
by introducing a set of operators $\{\vc{T},\vc{A}_{+},\vc{A}_{-}\}$
\begin{subequations}
\label{eq:17}
\begin{eqnarray}
\vc{A}_{+}(\bs{\theta},\bs{\theta^{\prime}})&=&
(\bar{\theta}-\bar{\theta}^{\prime})\theta,
\label{eq:17a}\\
\vc{A}_{-}(\bs{\theta},\bs{\theta^{\prime}})&=&
\vc{A}_{+}(\bs{\theta^{\prime}},\bs{\theta})=
-(\bar{\theta}-\bar{\theta}^{\prime})\theta^{\prime},
\label{eq:17b}\\
\vc{T}(\bs{\theta},\bs{\theta^{\prime}})&=&1,
\label{eq:17c}
\end{eqnarray}
\end{subequations}
equipped with the operator product
\begin{equation}
\mathbf{A}_{1}\cdot\mathbf{A}_{2}(\bs{\theta},
\bs{\theta^{\prime}})=
\int\text{d}^{2}\bs{\theta^{\prime\prime}}\,
\mathbf{A}_{1}(\bs{\theta},\bs{\theta^{\prime\prime}})
\mathbf{A}_{2}(\bs{\theta^{\prime\prime}},
\bs{\theta^{\prime}})\, ,
\label{eq:18}
\end{equation}
so that
\begin{equation}
\vc{G}=C \vc{T}+
G_{+}\vc{A}_{+}+
G_{-}\vc{A}_{-}\, .
\label{eq:19}
\end{equation}
 
It is not difficult to verify that the above operators form
the basis of algebra with respect to the operator product
\begin{subequations}
\label{eq:20}
\begin{eqnarray}
& &\mathbf{A}_{\pm}\cdot\mathbf{A}_{\pm}=\mathbf{A}_{\pm},\quad
\mathbf{T}\cdot\mathbf{A}_{+}=\mathbf{A}_{-}\cdot\mathbf{T}=\mathbf{T},
\label{eq:20a}\\
& &\mathbf{A}_{\pm}\cdot\mathbf{A}_{\mp}=
\mathbf{A}_{+}\cdot\mathbf{T}=\mathbf{T}\cdot\mathbf{A}_{-}=
\mathbf{T}\cdot\mathbf{T}=
\mathbf{0}.
\label{eq:20b}
\end{eqnarray}
\end{subequations}

In particular, Eqs.~(\ref{eq:20}) ease  finding of inversion formulae for
superoperators. For example, we can derive the expression for 
$\mathbf{G^{-1}}$:
\begin{equation}
\mathbf{G^{-1}}=G_{+}^{-1}\mathbf{A}_{+}+G_{-}^{-1}\mathbf{A}_{-}-
G_{-}^{-1}\cdot C\cdot G_{+}^{-1}\mathbf{T}\, ,
\label{eq:21}
\end{equation}
so that
\begin{equation}
\mathbf{G}\cdot\mathbf{G}^{-1}=
\delta(\bs{\theta}-\bs{\theta^{\prime}})\equiv
(\bar{\theta}-\bar{\theta}^{\prime})
(\theta-\theta^{\prime})=\vc{A}_{+}+\vc{A}_{-}\, .
\label{eq:22}
\end{equation}
Note that expansion of the bare correlation function
$\mathbf{G}^{(0)}$ over the above basis is
\begin{eqnarray}
\vc{G}^{(0)}&=&
\{
D^{(2)}+m
\}
\delta(\bs{\theta}-\bs{\theta^{\prime}})\delta_{ij}=
\nonumber\\
&=&\{-2\,\vc{T}+(m-\partial_{t})\vc{A}_{+}+
(m+\partial_{t})\vc{A}_{-}
\}\delta_{ij} .
  \label{eq:23}
\end{eqnarray}

The second identity Eq.~(\ref{eq:14}) provides the relation between the
autocorrelator of the order parameter $C(t,t^{\prime})$ 
and the response functions
$G_{-}(t,t^{\prime})\equiv G(t,t^{\prime})$ and
$G_{+}(t,t^{\prime})=G(t^{\prime},t)$
(retarded and advanced Green functions) known as
fluctuation--dissipation theorem (FDT):
\begin{subequations}
\label{eq:24}
\begin{eqnarray}
\frac{\partial}{\partial t}C(t,t^{\prime})&=&
-\theta(t-t^{\prime})G(t,t^{\prime})+
\theta(t^{\prime}-t)G(t^{\prime},t)
\label{eq:24a}\\
\frac{\partial}{\partial t^{\prime}}C(t,t^{\prime})&=&
\theta(t-t^{\prime})G(t,t^{\prime})-
\theta(t^{\prime}-t)G(t^{\prime},t)
\label{eq:24b}
\end{eqnarray}
\end{subequations}
where $\theta(t)$ is the step function.

It is convenient to reformulate 
FDT (\ref{eq:24})
in terms of time--dependent susceptibility $\chi(t,t^{\prime})$
\begin{equation}
  \label{eq:25}
\chi(t,t^{\prime})=
\int_{t^{\prime}}^{t}{\rm d}\tau^{\prime}G(t,\tau^{\prime})
\end{equation}
that gives the response to an applied field held constant from
time $t^{\prime}$ up to $t$. (Hereafter it is assumed that
$t\ge t^{\prime}$.) Since FDT implies the time translational invariance,
$G$, $C$ and $\chi$ depend only on the time separation 
$\tau\equiv t-t^{\prime}$. Then integrating Eq.~(\ref{eq:24b})
and taking the limit $\tau\to\infty$ yields the required form of FDT:
\begin{equation}
  \label{eq:26}
\chi=\int_{0}^{\infty}{\rm d}\tau^{\prime}\,G(\tau^{\prime})=q_0-q_h\, ,
\end{equation}
where $\chi$ is the static susceptibility,
$q_0=C(t,t)=C(0)$ and $q_h=C(\infty)$.

Analogously,  Ward identities 
for proper vertices lead to  the  same relations
for the mass operator (self--energy),
\begin{equation}
\bs\Sigma(z_1,z_2)=\Sigma(t_1,t_2)\vc{T}+
\Sigma_{+}(t_1,t_2)\vc{A}_{+}+
\Sigma_{-}(t_1,t_2)\vc{A}_{-}\, ,
\label{eq:27}
\end{equation}
that enter the Dyson equation. 

\section{Dynamical action for system with quenched disorder:
second Legendre transformation technique}
\label{sec:2}

In this section we derive the dynamical action as
a second Legendre transformation of the free energy
functional.
Since it is straightforward to generalize
the subsequent considerations, 
for the sake of simplicity, the technique will be employed
to study the system 
with the thermodynamic potential $V(\phi)$ 
of the following form
\begin{subequations}
  \label{eq:28}
\begin{eqnarray}
V(\phi)&=&\sum_{i}\{
U(\phi_{i}(z))+H_{i}\phi_{i}(z)
\} - \sum_{ij} W_{ij}\phi_{i}(z)\phi_{j}(z),
\label{eq:28a}\\
U(\phi)&=&\frac{m}{2!}\phi^2 + U_{anh}(\phi)=
\frac{m}{2!}\phi^2+
\frac{\lambda}{4!}\phi^4\, ,
\label{eq:28b}
\end{eqnarray}
\end{subequations}
where $m=\mu T$.
Under suitable assumptions, it can be regarded as a discrete version
of the well--known Landau--Ginzburg free energy functional for 
coarse--grained scalar order parameter field~\cite{Bray}.
We consider the case where the couplings $W_{ij}$ and 
the field $H_i$ correspond
to quenched degrees of freedom and are independent Gaussian variables:
$$
\overline{W}_{mn} =0,\quad
\overline{W_{mn}W}_{m^{\prime}n^{\prime}}=
\delta_{mm^{\prime}} \delta_{nn^{\prime}}w_{mn},\quad
\overline{H}_{m} =0,\quad
\overline{H_{m}H}_{m^{\prime}}=
\delta_{mm^{\prime}} h^2\, . 
$$

Averaging away the quenched variables gives
\begin{subequations}
  \label{eq:29}
\begin{eqnarray}
\overline{\exp\{-S\}}&=&\exp\{-T^{-1}
\int{\rm d}z{\rm d}z^{\prime}{ L}_{eff}\}
\label{eq:29a}\\
{ L}_{eff}&=&\sum_{i}\left\{
\frac{1}{2}\phi_{i}(z)K(z,z^{\prime})\phi_{i}(z^{\prime})+
\delta(z-z^{\prime})U_{anh}(\phi_{i}(z))\right\}-
\nonumber \\
& &-(2T)^{-1}\sum_{ij}
w_{ij}\phi_{i}(z)\phi_{j}(z)\phi_{i}(z^{\prime})
\phi_{j}(z^{\prime})  
\end{eqnarray}
\end{subequations}
where (see Eq.~(\ref{eq:23}))
\begin{equation}
  \label{eq:30}
  K(z,z^{\prime})=
\left[
-\frac{h^{2}}{T}\vc{T}
+\left(2\vc{T}+(m-\partial_{t})\vc{A}_{+}+
(m+\partial_{t})\vc{A}_{-}\right)\delta(t-t^{\prime})
\right]\delta_{ij}
\,. 
\end{equation}
Note that, by contrast with the real two--particle interaction,
the term generated by the averaging is non--local in time. 

At this stage we can use functional methods of the field theory 
to derive equations of motion for
the one-- and two--point correlators,
$\langle\phi_{i}(z)\rangle$ and
$\langle\phi_{i}(z)\phi_{j}(z^{\prime})\rangle$.
To this end let us write the effective action (\ref{eq:29})
in slightly generalized form:
\begin{equation}
  \label{eq:31}
  S_{eff}=\sum_i S_0(\phi_i|A^{(i)})+\frac{1}{2}\int{\rm d}z_1{\rm d}z_2
\sum_{ij}\bar{w}_{ij}\phi_i(z_1)\phi_i(z_2)\phi_j(z_1)\phi_j(z_2)\, ,
\end{equation}
where
\begin{equation}
  \label{eq:32}
  S_0(\phi_i|A^{(i)})=
\sum_{n=1}^{4}\frac{1}{n!}\int{\rm d}z_1\ldots{\rm d}z_n
A_n^{(i)}(z_1,\ldots z_n)\phi_i(z_1)\ldots\phi_i(z_n)
\end{equation}
and $A^{(i)}\equiv\{A_1^{(i)},A_2^{(i)},A_3^{(i)},A_4^{(i)}\}$
stands for a set of "superpotentials" that define the one--particle
action $S_0$ and are assumed to be symmetric functions of
arguments. In our case $A_1^{(i)}=A_3^{(i)}=0$,
$A_2^{(i)}=-K(z_1,z_2)/T$ and
$A_4^{(i)}=-\lambda/T\delta(z_1-z_2)\delta(z_2-z_3)\delta(z_3-z_4)$.
For convenience,
we will keep the generalized denotions and the above
$A_k^{(i)}$ will be inserted into the final equations.
Note that
$A_1$ is commonly referred to as an external field and
the inverse of operator with
the kernel $-A_2(z_1,z_2)$  
is proportional to the bare correlation function.
 
The effective action $S_{eff}$ defines the partition function
$Z(A)$ and the free energy $F(A)$ as generating functionals
of $n$--point correlators without vacuum loops and
connected $n$--point correlators, respectively:
\begin{subequations}
  \label{eq:33}
\begin{eqnarray}  
Z(A)=\exp\{F(A)\}&=&N^{-1}\int\prod_i D\phi_i\exp\{S_{eff}(\phi|A)\}\, ,
\label{eq:33a}\\
\frac{\delta^{n} Z(A)}{\delta A_1^{(i)}(z_1)\ldots\delta A_1^{(i)}(z_n)}
&=&\langle\phi_i(z_1)\ldots\phi_i(z_n)\rangle\equiv n!\, \alpha_n^{(i)}
(z_1\ldots z_n)\, ,
\label{eq:33b}\\
\frac{\delta^{n} F(A)}{\delta A_1^{(i)}(z_1)\ldots\delta A_1^{(i)}(z_n)}
&=&\langle\phi_i(z_1)\ldots\phi_i(z_n)\rangle_{c}\equiv \beta_n^{(i)}
(z_1\ldots z_n)\, ,
\label{eq:33c}\\
\frac{\delta F(A)}{\delta A_2^{(i)}(z_1,z_2)}
&=&\alpha_2^{(i)}(z_1,z_2)\, .
\label{eq:33d}
\end{eqnarray}
\end{subequations}

After introducing auxiliary field $X_i(z_1,z_2)$ 
and performing the Hubbard--Stratonovich transformation
\begin{eqnarray}
  \label{eq:34}
& &\exp\left[
\frac{1}{2}\int{\rm d}z_1{\rm d}z_2
\sum_{ij}\bar{w}_{ij}\phi_i(z_1)\phi_i(z_2)\phi_j(z_1)\phi_j(z_2)
\right]\propto
\nonumber\\
& &\phantom{\exp[-\frac{1}{2}}
\int\prod_i DX_i
\exp \Big[
-\frac{1}{2}\int{\rm d}z_1{\rm d}z_2
\sum_{ij}\bar{w}_{ij}^{-1}X_i(z_1,z_2)X_j(z_1,z_2)+
\nonumber\\
& &\phantom{\exp[-\frac{1}{2}} +
\sum_i X_i(z_1,z_2)\phi_i(z_1)\phi_i(z_2)
 \Big]
\end{eqnarray}
the partition function assumes the following form
\begin{eqnarray}
  \label{eq:35}
Z(A)&=&
\int\prod_i DX_i
\exp \Big[
-\frac{1}{2}\int{\rm d}z_1{\rm d}z_2
\sum_{ij}\bar{w}_{ij}^{-1}X_i(z_1,z_2)X_j(z_1,z_2)+
\nonumber\\
& &\phantom{\exp[-\frac{1}{2}} +
\sum_i F_0(A_1^{(i)},A_2^{(i)}+2 X_i,A_3^{(i)},A_4^{(i)})
 \Big]\, ,
\end{eqnarray}
where $F_0(A^{(i)})$ is the generating functional of
connected correlators for the one--particle action $S_0$~:
 \begin{equation}
  \label{eq:36}
\exp\{F_0(A^{(i)})\}=
\int D\phi_i \exp\{S_0(\phi_i|A^{(i)})\}\, .
\end{equation}

After changing variables 
${\tilde{A}}_{2}^{(i)}=A_2^{(i)}+2 X_i$ in
Eq.~(\ref{eq:35}) it is ready to derive the equations
of the saddle--point approximation:
\begin{equation}
  \label{eq:37s}
  \frac{\delta F_0}{\delta \tilde{A}_{2}^{(i)}}=
\alpha_{2}^{(i)}=\frac{1}{4}
\sum_j \bar{w}_{ij}^{-1}(\tilde{A}_{2}^{(i)}-A_{2}^{(i)})\, .
\end{equation}

It is known that the mean field approximation
is given by the leading order in steepest descent
calculations of such kind~\cite{Justin}. The only difference
from the standard mean field theory
is that the interaction term in Eq.~(\ref{eq:35})
contains ${A}_{2}^{(i)}$ instead of ${A}_{1}^{(i)}$.
So it is natural to 
define second Legendre transformation
for $F$ as a generalized Legendre transformation
with respect to the first two "superpotentials"
$A_1$ and $A_2$~\cite{Vasil'ev}~:
\begin{eqnarray}
\Gamma(\alpha_1,\alpha_2|A_3,A_4)&=&
F(A)-A_1^{(i)}\frac{\delta F}{\delta A_1^{(i)}}-
A_2^{(i)}\frac{\delta F}{\delta A_2^{(i)}}=
\nonumber\\
&=&F(A)-A_1\alpha_1-
A_2\alpha_2\, ,
  \label{eq:37}
\end{eqnarray}
where 
$$
\alpha_1^{(i)}(z_1)=\beta_1^{(i)}(z_1)=\langle\phi_i(z_1)\rangle\, ,
\quad  
2\alpha_2^{(i)}(z_1,z_1)=\beta_2^{(i)}(z_1,z_2)+
\beta_1^{(i)}(z_1)\beta_1^{(i)}(z_2)
$$
and universal functional
notations used in (\ref{eq:37}) imply that
$$
A_1\alpha_1\equiv\sum_i
\int{\rm d}z_1 A_1^{(i)}(z_1)\alpha_1^{(i)}(z_1)\, , \quad
A_2\alpha_2\equiv\sum_i
\int{\rm d}z_1{\rm d}z_2 A_2^{(i)}(z_1,z_2)\alpha_2^{(i)}(z_1,z_2)\, .
$$
Similar to the standard Legendre transformation,
given the functional $\Gamma$ with
$A_1$ and $A_2$, the correlators $\alpha_1$ and $\alpha_2$
are defined as solutions to the equations
\begin{subequations}
  \label{eq:39}
\begin{eqnarray}
& &\frac{\delta\Gamma}{\delta\alpha_1^{(i)}(z_1)}=-A_1^{(i)}(z_1)\\
& &\frac{\delta\Gamma}{\delta\alpha_2^{(i)}(z_1,z_2)}=
-A_2^{(i)}(z_1,z_2)\, . 
\end{eqnarray}
\end{subequations}
Eqs.~(\ref{eq:39}) are Euler--Lagrange equations for the functional
\begin{equation}
  \label{eq:40}
  \Phi(\alpha_1,\alpha_2|A)=
\Gamma(\alpha_1,\alpha_2|A_3,A_4)+\alpha_1 A_1+\alpha_2 A_2\, ,
\end{equation}
so that substituting of extremals from Eqs.~(\ref{eq:39}) into $\Phi$ 
provides the corresponding values of the free energy $F$.
It follows that $\Phi$ has the meaning of dynamical action.

In the mean field approximation,
from Eq.~(\ref{eq:37s}) the expression for $\Gamma$ is
\begin{equation}
  \label{eq:41}
\Gamma = \sum_i \Gamma_0(\alpha_1^{(i)},\alpha_2^{(i)}|A_3,A_4)+  
2 \sum_{ij} \bar{w}_{ij}\int{\rm d}z_1{\rm d}z_2 
\alpha_2^{(i)}(z_1,z_2)\alpha_2^{(j)}(z_1,z_2)\, ,
\end{equation}
where $\Gamma_0$ is the Legendre transform of the one--particle
free energy. As a functional of $\beta_1$ and $\beta_2$, $\Gamma_0$
is given by~\cite{Vasil'ev}
\begin{equation}
  \label{eq:42}
  \Gamma_0=\frac{1}{2}\tr\ln\beta_2+\bar{\Gamma}_0\, .
\end{equation}
where $\bar{\Gamma}_0$ can be calculated as a sum of
two--particle irreducible diagrams (it cannot be made
disconnected by cutting off two  lines)
with "dressed" internal lines ($\beta_2$ is the propagator).
Up to the second order of the perturbation theory with
$A_3=0$, we have
\begin{equation}
\label{eq:43}
\bar{\Gamma}_0=
\frac{1}{4!}A_4\,\left(
\beta_1^{4}+6\beta_2\beta_1^{2}+3\beta_2^{2}
\right)+
\frac{1}{48}A_4\beta_2^{4}A_4+
\frac{1}{12}(A_4\beta_1)\beta_2^{3}(A_4\beta_1) .
\end{equation}
The diagrammatic representation of the terms on 
the right hand side of Eq.~(\ref{eq:43}) is
\begin{fmffile}{diagram}
\begin{eqnarray}
\frac{1}{24}\quad
\parbox{20mm}{
\begin{fmfgraph}(15,10)
\fmfpen{thick}
\fmfleft{i1,i2} \fmfright{o1,o2}
\fmf{zigzag}{v,i1} \fmf{zigzag}{v,i2}
\fmf{zigzag}{v,o1} \fmf{zigzag}{v,o2}
\fmfdot{v} 
\end{fmfgraph}
} 
&+&\quad \frac{1}{4}\parbox{20mm}{
\begin{fmfgraph}(20,15)
\fmfpen{thick}
\fmfleft{i} \fmfright{o1,o2}
\fmf{phantom}{i,v1} 
\fmf{zigzag}{v2,o1} \fmf{zigzag}{v2,o2}
\fmf{phantom,left=.5,tension=0.3}{v1,v2,v1}\fmffreeze
\fmf{plain,left}{v1,v2,v1}
\fmfdot{v2} 
\end{fmfgraph}
}
+\quad \frac{1}{8}\parbox{20mm}{
\begin{fmfgraph}(20,15)
\fmfpen{thick}
\fmfleft{i} \fmfright{o}
\fmf{phantom}{i,v}
\fmf{phantom}{v,o}
\fmf{plain}{v,v} 
\fmf{plain,left=90}{v,v}
\fmfdot{v} 
\end{fmfgraph}
} +\nonumber\\
&+&\quad \frac{1}{48}\parbox{20mm}{
\begin{fmfgraph}(20,15)
\fmfpen{thick}
\fmfleft{i} \fmfright{o}
\fmf{phantom}{i,v1} 
\fmf{phantom}{v2,o}
\fmf{plain,left=.5,tension=0.3}{v1,v2,v1}\fmffreeze
\fmf{plain,left}{v1,v2,v1}
\fmfdotn{v}{2} 
\end{fmfgraph}
} 
+\quad \frac{1}{12}\quad\parbox{20mm}{
\begin{fmfgraph}(20,15)
\fmfpen{thick}
\fmfleft{i} \fmfright{o}
\fmf{zigzag}{i,v1} 
\fmf{zigzag}{v2,o}
\fmf{plain,straight}{v1,v2}
\fmffreeze
\fmf{plain,left}{v1,v2,v1}
\fmfdotn{v}{2} 
\end{fmfgraph}
}\quad .\nonumber
\end{eqnarray}
\end{fmffile}
In the diagrams zigzag lines correspond to
$\beta_1$--lines directly joined to the vertices. 
The equations of motion in new variables are
\begin{subequations}
  \label{eq:44}
\begin{eqnarray}
& &\frac{\delta\Gamma}{\delta\beta_1(z_1)}
-2\frac{\delta\Gamma}{\delta\beta_2(z_1,z_2)}\,\beta_1(z_2)
=-A_1(z_1)
\label{eq:44a}\\
& &2\frac{\delta\Gamma}{\delta\beta_2(z_1,z_2)}=
-A_2(z_1,z_2)\, .
\label{eq:44b} 
\end{eqnarray}
\end{subequations}
Note that, owing to the identity
$$
\frac{\delta}{\delta\beta_2(z_1,z_2)}\tr\ln\beta_2=\beta_2^{-1}(z_1,z_2)\, ,
$$
Eq.~(\ref{eq:44b}) is the Dyson equation with 
suitably defined  mass operator.

In what follows it is supposed that the system under consideration is 
spatially homogeneous, so that the correlators
$\beta_1(z)\equiv\langle\phi\rangle$ and
$\beta_2(z,z^{\prime})\equiv\vc{G}(z,z^{\prime})$ do not depend on the
site index. Moreover, since Eq.~(\ref{eq:44a}) has a trivial solution
$\langle\phi\rangle=0$ at $A_1=0$ and ergodicity breaking transition
is related to the anomaly in $\vc{G}$ when FDT is violated, we eliminate
$\langle\phi\rangle$ from the consideration by putting
$\langle\phi\rangle=0$.
Eqs.~(\ref{eq:40}-\ref{eq:43}) then provide the expression
for the dynamical action:
\begin{eqnarray}
  \label{eq:45}
2\Phi&=& \tr\ln[\vc{G}]+
\frac{\lambda^2}{24 T^2}\int{\rm d}z_1{\rm d}z_2
\vc{G}^4(z_1,z_2)+\nonumber\\
&+&T^{-1}\int{\rm d}z_1{\rm d}z_2
\left[\,\frac{w}{2T}\,\vc{G}^2(z_1,z_2)-
K(z_1,z_2)\vc{G}(z_1,z_2)\,
\right]  
\end{eqnarray}
where $w=\sum_{i} w_{ij}$. 
The corresponding Dyson equation follows from 
the stationarity condition
$\displaystyle
\frac{\delta\Phi}{\delta\vc{G}(z,z^{\prime})}=0$ :
\begin{equation}
  \label{eq:46}
\vc{G}^{-1}(z,z^{\prime})=\frac{1}{T}
K(z,z^{\prime})-\frac{w}{T^{2}}\vc{G}(z,z^{\prime})
-\bs{\Sigma}\,(z,z^{\prime})\, .
\end{equation}
 
The mass operator 
$
\bs{\Sigma}\,(z,z^{\prime})=
\lambda^2/(6T^{2})\vc{G}^{3}(z,z^{\prime})
$ 
has the following components
\begin{equation}
  \label{eq:47}
\Sigma(t,t^{\prime})=\Sigma(C)=
\frac{\lambda^2}{6T^{2}}C^{3}(t,t^{\prime})\, ,
\qquad
\Sigma_{\pm}(t,t^{\prime})=\Sigma^{\, \prime}(C)\,G_{\pm}\, ,
\end{equation}
where
$\displaystyle
\Sigma^{\, \prime}=\frac{\partial\Sigma}{\partial C} $.
Clearly, $\bs{\Sigma}$ satisfies the FDT:
$\partial_{t^{\prime}}\Sigma=
-\partial_{t}\Sigma=\Sigma_{-}$.

By using the identity (\ref{eq:21}), it is not difficult to derive
the components of Eq.~(\ref{eq:46}).
So the Dyson equation  can be represented as a system
of nonlinear integro--differential equations:
\begin{equation}
  \label{eq:48}
G_{\pm}^{-1}=\frac{1}{T}\delta(t-t^{\prime})(m \mp \partial_t)-
\frac{w}{T^2}G_{\pm}-\Sigma_{\pm}\, ,
\end{equation}
\begin{equation}
  \label{eq:49}
G_{-}^{-1}\cdot C\cdot G_{+}^{-1}=\frac{2}{T}\delta(t-t^{\prime})+
\frac{h^2}{T^2}+\frac{w}{T^2} C +\Sigma\, .
\end{equation}
Recall that the dot "$\cdot$" denotes the operator product.

Finally, we arrive at the explicit form of the dynamical
Dyson equations for the autocorrelator $C(t,t^{\prime})$
and the response (Green) function $G(t,t^{\prime})$:
\begin{eqnarray}
\label{eq:50}
\frac{1}{T}(m+\partial_{t})G(t,t^{\prime})&=&
\delta(t-t^{\prime})+
\frac{w}{T^2}\int_{t^{\prime}}^{t}{\rm d}\tau^{\prime}\,
G(t,\tau^{\prime})G(\tau^{\prime},t^{\prime})+
\nonumber\\
&+&\int_{t^{\prime}}^{t}{\rm d}\tau^{\prime}\,
\Sigma_{-}(t,\tau^{\prime})G(\tau^{\prime},t^{\prime})
\end{eqnarray}
\begin{eqnarray}
\label{eq:51}
\frac{1}{T}(m&+&\partial_{t})C(t,t^{\prime})=
\frac{2}{T}G(t^{\prime},t)+\frac{h^2}{T^2}
\int_{-\infty}^{t^{\prime}}{\rm d}\tau^{\prime}\,G(t^{\prime},t)+
\nonumber\\
&+&\frac{w}{T^2}
\left[
\int_{-\infty}^{t^{\prime}}{\rm d}\tau^{\prime}\,
C(t,\tau^{\prime})G(t^{\prime},\tau^{\prime})
+\int_{-\infty}^{t}{\rm d}\tau^{\prime}\,
G(t,\tau^{\prime})C(\tau^{\prime},t^{\prime})
\right]+
\nonumber\\
&+&
\int_{-\infty}^{t^{\prime}}{\rm d}\tau^{\prime}\,
\Sigma(t,\tau^{\prime})G(t^{\prime},\tau^{\prime})
+\int_{-\infty}^{t}{\rm d}\tau^{\prime}\,
\Sigma_{-}(t,\tau^{\prime})C(\tau^{\prime},t^{\prime})
\end{eqnarray}

\section{Ergodicity breaking transition}
\label{sec:4}
In this section we concentrate on asymptotic analysis
of the Dyson equations (\ref{eq:50},\ref{eq:51}). Our primary
purpose is to describe the behavior of the system in terms of
static susceptibilities and memory parameters characterizing
asymptotics of the autocorrelator $C(t,t^{\prime})$ in the
limit of large time separation, $\tau=t-t^{\prime}\to\infty$.

The first step is to transform Eq.~(\ref{eq:51}) in such a way
that FDT is not broken explicitly. To this end we assume
that interaction is adiabatically switched ($C(-\infty)$=0)
and apply FDT on the interval of time from $-\infty$ to $t^{\prime}$.
The result reads
\begin{eqnarray}
\label{eq:52}
\frac{1}{T}(m&+&\partial_{t})C(t,t^{\prime})=
q_0\left\{
\frac{h^2}{T^2}+
\frac{w}{T^2}C(t,t^{\prime})+
\Sigma(t,t^{\prime})
\right\}+
\nonumber\\
&+&\frac{w}{T^2}
\int_{t^{\prime}}^{t}{\rm d}\tau^{\prime}\,
G(t,\tau^{\prime})C(\tau^{\prime},t^{\prime})
+\int_{t^{\prime}}^{t}{\rm d}\tau^{\prime}\,
\Sigma_{-}(t,\tau^{\prime})C(\tau^{\prime},t^{\prime})\, ,
\end{eqnarray}
where $q_0$ is the value of equal time autocorrelator 
$C(t^{\prime},t^{\prime})$.
Thus the starting point for subsequent analysis
is the Dyson equations (\ref{eq:50}) and (\ref{eq:52}).

\subsection{High temperature region}
Since in this region the time translational invariance
is unbroken, Eqs.~(\ref{eq:50},\ref{eq:52}) can be easily
analyzed by making use of Laplace transformation.
The resulting system for the Laplace transforms
$\hat{C}(p)$ and $\hat{G}(p)$ is given by
\begin{subequations}
\label{eq:53}
\begin{eqnarray}
\frac{1}{T}(m+p)\hat{G}(p)&=&
1+\frac{w}{T^2}\hat{G}^{2}(p)+\hat{\Sigma}_{-}(p)\hat{G}(p)\, ,
\label{eq:53a}\\
\frac{1}{T}[(m+p)\hat{C}(p)-q_0]&=&
q_0\left(
\frac{h^2}{pT^2}+\frac{w}{T^2}\hat{C}(p)+\hat{\Sigma}(p)
\right)+\nonumber\\
&+&\hat{C}(p)\left(
\frac{w}{T^2}\hat{G}(p)+\hat{\Sigma}_{-}(p)
\right)\, .
\label{eq:53b}
\end{eqnarray}
\end{subequations}
By using the relations: $\chi=\hat{G}(0)$, 
$q_0=\lim_{p\to\infty} p\hat{C}(p)$, $q_h=\lim_{p\to 0}p\hat{C}(p)$
and $\Sigma(q_h)=\lim_{p\to 0}p\hat{\Sigma}(p)$, from
Eq.~(\ref{eq:53a}) and Eq.~(\ref{eq:53b}) multiplied by $p$
the following system for susceptibilities and the field induced
parameter can be deduced
\begin{subequations}
\label{eq:54}
\begin{eqnarray}
\frac{m}{T}\chi&=&
1+\frac{w}{T^2}\chi^2+\hat{\Sigma}_{-}(0)\chi\, ,
\label{eq:54a}\\
\frac{m}{T}\chi_{a}&=&
1+\frac{w}{T^2}\chi_a^2+\chi_a\left\{\Sigma(\chi_a+q_h)-\Sigma(q_h)\right\}
-\nonumber\\
&-&q_h\left\{\frac{w}{T^2}\Delta\chi+\Delta\hat{\Sigma}_{-}(0)\right\}\, ,
\label{eq:54b}\\
q_h&=&\chi(\chi_a+q_h)
\left\{
\frac{h^2}{T^2}+\frac{w}{T^2} q_h +\Sigma(q_h)
\right\}\, ,
\label{eq:54c}
\end{eqnarray}
\end{subequations}
where $\chi_a=q_0-q_h$,  
$\Delta\chi=\chi-\chi_a$ and
$\Delta\hat{\Sigma}_{-}(0)=
\hat{\Sigma}_{-}(0)+\Sigma(q_h)-\Sigma(\chi_a+q_h)$.
It should be stressed that, except for the identity
$\lim_{p\to\infty}p(p\hat{C}(p)-q_0)
=-\lim_{p\to\infty}p\hat{G}(p)=-T $
that yield the first term on the right hand side
of Eq.~(\ref{eq:54b}), FDT have not been used in the derivation
of the above equations.

In the FDT regime $\Delta\chi=0$ and $\Delta\hat{\Sigma}_{-}(0)=0$.
So the equations for the susceptibility $\chi$ and
the field induced parameter $q_h$ in the high temperature region
are
\begin{subequations}
\label{eq:55}
\begin{eqnarray}
\frac{m}{T}\chi&=&
1+\frac{w}{T^2}\chi^2+\chi\left\{\Sigma(\chi+q_h)-
\Sigma(q_h)\right\}\, ,
\label{eq:55a}\\
q_h&=&\chi(\chi+q_h)
\left\{
\frac{h^2}{T^2}+\frac{w}{T^2} q_h +\Sigma(q_h)
\right\}\, .
\label{eq:55b}
\end{eqnarray}
\end{subequations}
Notice that, in the case of $h^2=0$, the temperature dependence
of $\chi$ is defined by Eq.~(\ref{eq:55a}) with $q_h=0$.
This is why the parameter $q_h$ is referred to as a field induced
parameter (there are no memory effects in the absence of random
field at $T>T_c$).

\subsection{Critical point}
Ergodicity breaking transition is determined by the point where
the FDT compliant solution with $C_{\rm FDT}(t)$ and 
$G_{\rm FDT}(t)$ becomes dynamically unstable.
In order to study stability let us first differentiate Eq.~(\ref{eq:52})
with respect to time and then insert the perturbed solution
$\partial_t C=\partial_tC_{\rm FDT}+\delta D$ 
and $G=G_{\rm FDT}+\delta G$ into the resulting system.
As far as stability analysis is concerned only linear part
of the system is relevant. 
After rather straightforward calculations, it
can be derived in the form:
\begin{subequations}
\label{eq:56}
\begin{eqnarray}
\left[
\frac{1}{T}\,\partial_t+\chi^{-1}
-\chi\left(\frac{w}{T^2}+\Sigma^{\prime}(q_h)\right)
\right]\delta G&=&\ldots \, ,
\label{eq:56a}\\
\left[
\frac{1}{T}\,\partial_t+\chi^{-1}
-(\chi+q_h) \left(\frac{w}{T^2}+\Sigma^{\prime}(q_h)\right)
\right]\delta D &-&
\nonumber\\
-q_h \left(\frac{w}{T^2}+\Sigma^{\prime}(q_h)\right)
\delta G&=&\ldots \, ,
\label{eq:56b}
\end{eqnarray}
\end{subequations}  
where "\ldots" stands for the terms that are
nonlinear in $\delta D$ and $\delta G$.

As a result, the marginal stability condition reads
\begin{equation}
  \label{eq:57}
1=\chi(\chi + q_h)
\left\{\frac{w}{T^2}+\Sigma^{\prime}(q_h)\right\}\, .
\end{equation}
It should be emphasized that fluctuations do not need to
obey FDT, $\delta D+\delta G\ne 0$. As a
consequence, at $q_h\ne 0$ Eq.~(\ref{eq:57}) is stronger than
the condition 
\begin{equation}
  \label{eq:57a}
1=\chi^2
\left\{\frac{w}{T^2}+\Sigma^{\prime}(q_h)\right\}\, .
\end{equation}
obtained under the assumption that fluctuations do not violate
FDT ($\delta D+\delta G= 0$).

Eqs.~(\ref{eq:55},\ref{eq:57}) yield the temperature of ergodicity
breaking transition, $T_c$, and the value of $q_h$  at the
critical point, $q_c$. The latter can be easily obtained from
Eqs.~(\ref{eq:55b},\ref{eq:57}): $q_c=(3h^2/\lambda^2)^{1/3}$.

\subsection{Behavior at the limit $T\to T_c-0$}

Despite the detailed analysis of low temperature region is beyond the
scope of this paper it is instructive to see how equation that define
the limiting value of susceptibility when approaching $T_c$ from
below is related to the marginal stability condition.

Let us first consider Eqs.~(\ref{eq:54})
and suppose that near $T_c$ FDT is modified as follows:
\begin{equation}
  \label{eq:58}
-\frac{{\rm d}}{{\rm d}t}\,C(t)=G(t)-\Delta G(t)\, ,
\qquad
\Delta G(t) = - \frac{{\rm d}}{{\rm d}t}\Psi(C(t))\, ,
\end{equation}
where  $\Psi$ is a nondecreasing function
which derivative $\Psi^{\prime}$ vanishes outside the interval
$[q_1,q_2]$ with $q_2<q_0$. Obviously, these assumptions
ensure the validity of Eq.~(\ref{eq:54}) and lead to the following
results:
\begin{equation}
  \label{eq:59}
  \Delta\chi = \Psi(q_2)-\Psi(q_1)\, ,
\end{equation}
\begin{equation}
  \label{eq:60}
  \Delta\hat{\Sigma}_{-}(0) = 
\int_{q_1}^{q_2}{\rm d}q\,\Sigma^{\prime}(q)\Psi^{\prime}(q)=
\Sigma^{\prime}(q_m)\Delta\chi\, ,
\end{equation}
where $q_m\in (q_1,q_2)$ is the middle point.

Below $T_c$ $\Delta\chi\ne 0$, so subtracting Eq.~(\ref{eq:54b}) 
from Eq.~(\ref{eq:54a}) and dividing the result by $\Delta\chi$
gives
\begin{equation}
  \label{eq:61}
  1=\chi(\chi + q_h)
\left\{\frac{w}{T^2}+\Sigma^{\prime}(q_m)\right\}\, ,
\end{equation}
where the terms proportional to $\Delta\chi$ are neglected.
From Eqs.~(\ref{eq:54c},\ref{eq:61}) $q_m\to q_h$ at $T\to T_c-0$ and
we recover the marginal stability condition as equation
for the limiting value of susceptibilities.

Clearly, aging cannot be taken into account provided the
time translational invariance is unbroken, so
the above consideration is closely
related to the regime known as true ergodicity
breaking~\cite{Parisi2,Barrot}. 
In this case the system equilibrates in a separate ergodic
component. 
Alternatively, according to the concept of weak
ergodicity breaking~\cite{Bouchaud}, the system does not equilibrate
and asymptotically ($t,t^{\prime}\gg 1 $) 
FDT is violated at large separation times,
$\tau\sim t^{\prime}$, whereas the system reveals 
quasi--equilibrium behavior at sufficiently small $\tau$.
It implies that the generalized form of FDT breaking contributions
to the correlation functions is~\cite{Cugliandolo1}
\begin{equation}
  \label{eq:62}
  C(t,t^{\prime})=C_0(t-t^{\prime})+C_a(t^{\prime}/t),\quad
  G(t,t^{\prime})=G_0(t-t^{\prime})+t^{-1}G_a(t^{\prime}/t),
\end{equation}
and the modification of FDT is given by the following
relations
\begin{equation}
  \label{eq:63}
\partial_{\,t^{\prime}}C_0(t-t^{\prime})=G_0(t-t^{\prime}),\quad
  x \partial_z C_a(z)=G_a(z),
\end{equation}
where $ z\equiv t^{\prime}/t$ and $x$ parameterizes the violation
of FDT.
The time--dependent susceptibility $\chi(t,t^{\prime})$
(see Eq.~(\ref{eq:25})) now depends on both the waiting time $t^{\prime}$
and the separation time $\tau$. 
So the value of static susceptibility is different
depending on the order in which limits $t^{\prime}\to\infty$ and
$\tau\to\infty$ are taken
\begin{eqnarray}
  \label{eq:64}
\lim_{\tau\to\infty}\lim_{t^{\prime}\to\infty}
\chi(t^{\prime}+\tau,t^{\prime})&=&\chi_a=q_0-q,\nonumber\\
\lim_{\tau\to\infty}\chi(t^{\prime}+\tau,t^{\prime})&=&\chi
=\chi_a+x \Delta q\, ,
\end{eqnarray}
where $q_0\equiv C(t,t)=C_0(0)+C_a(1)$, 
$\Delta q\equiv q-q_h=C_a(1)-C_a(0)$ and
\begin{eqnarray}
  \label{eq:65}
q&=&
\lim_{\tau\to\infty}\lim_{t^{\prime}\to\infty}
C(t^{\prime}+\tau,t^{\prime})=C_0(\infty)+C_a(1)\, ,\nonumber\\
q_h&=&
\lim_{\tau\to\infty}C(t^{\prime}+\tau,t^{\prime})
=C_0(\infty)+C_a(0)\, .
\end{eqnarray}
In the sense of Cubo~\cite{Cubo}, $\chi_a$ and $\chi$
can be referred to as adiabatic (thermodynamic) and isothermal
susceptibilities, respectively; $q$ is the dynamical
Edwards--Anderson parameter. Inserting Eq.~(\ref{eq:62}) into
Eq.~(\ref{eq:50}) integrated over the second argument of $G$
gives the system
\begin{subequations}
\label{eq:66}
\begin{eqnarray}
\frac{m}{T}\chi_a&=&
1+\frac{w}{T^2}\chi_a^2+\chi_a\left\{\Sigma(\chi_a+q)-
\Sigma(q)\right\}\, ,
\label{eq:66a}\\
\frac{m}{T}\chi&=&
1+\frac{w}{T^2}\chi^2+\chi\left\{\Sigma(\chi_a+q)-
\Sigma(q_h)+(x-1)\Delta q\Sigma^{\prime}(q_h)
\right\}\, .
\label{eq:66b}
\end{eqnarray}
\end{subequations}
In exactly the same manner
as the marginal stability condition (\ref{eq:57})
was recovered from Eqs.~(\ref{eq:54},\ref{eq:59},\ref{eq:60}),
Eqs.~(\ref{eq:66}) yield the condition
(\ref{eq:57a}) in the limit $\Delta q\to 0$ at $T\to T_c-0$.
Interestingly, it follows that, despite the susceptibility
$\chi$
and non--ergodicity parameter $\Delta q$ are continuous
at the critical point, $q$ can be discontinuous.

So, in both cases $\Delta\chi$ plays the role of order parameter, but,
in general, equations for the limiting values of susceptibilities are
different depending on whether the aging is taken into consideration.
Assuming the time homogeneity implies that the system can be described
by Eqs.(\ref{eq:54}) below $T_c$, but it can be shown 
that in this case we would 
encounter discontinuity of the susceptibility
at the critical point. Thus we arrive at the conclusion that aging
plays an important part in the problem under consideration. 

\section{Numerical results and discussion}
\label{sec:5}

In the above section
the ergodicity breaking 
transition is treated on the basis of the dynamical approach.
The critical temperature is defined as the point 
of marginal stability for FDT complaint solution.
At $h^2=0$ Eqs.~(\ref{eq:55a},\ref{eq:57}) are easy to solve,
so the critical temperature $T_c$ is given by
\begin{equation}
  \label{eq:67}
  T_c=4\frac{w^{3/2}}{\mu w_c}
\left(1-\sqrt{1-w_c/w}\right)\, ,\quad
w_c=3\left[\frac{2\lambda}{3\mu}\right]^2\, .
\end{equation}
From Eq.~(\ref{eq:67}) it is clear that, 
in order for the transition to occur,
the disorder intensity $w$ must exceed its critical $w_c$.
Dependence of $w_c$ on the intensity of random field is depicted
in Fig.~1. It is seen that there is no threshold for $w$ at sufficiently
large intensities of random field, 
whereas $w_c$ is an increasing function of $h^2$ in the range of small
intensities.
Note that, in order to simplify analysis, the relevant quantities are
rescaled as follows:
$$
\chi\rightarrow\mu\chi ,\quad q\rightarrow\mu q ,\quad
w\rightarrow 6\mu^2\lambda^{-2} w ,\quad
h^2\rightarrow 6\mu^3\lambda^{-2} w ,\quad
T\rightarrow\sqrt{6}\mu^2\lambda^{-1} T\, ,
$$
so the system for $T_c$ and $\chi_c$ at the critical point
takes the form:
\begin{subequations}
\label{eq:68}
\begin{eqnarray}
(\chi_c-1)\,T_c^2&=&w\chi_c^2+\chi_c
\left[\,(\chi_c+q_c)^3-q_c^3 \,\right]\, ,
\label{eq:68a}\\
T_c^2&=&
\chi_c\,(\chi_c+q_c)
\left[\,w+3 q_c^2\,\right]\, ,
\label{eq:68b}
\end{eqnarray}
\end{subequations}
where $q_c=h^{2/3}/2$.
Dependencies of $T_c$ and the temperature of freezing transition,
where the derivative of $\chi$ with respect to temperature
diverges, $T_f$ on the disorder intensity $w$ at $h^2=0$
are shown in Fig.~2. As is seen, the difference $T_c-T_f$ 
goes to zero when  $w$ increases.
The latter is not very surprising, for $T_c=T_f$ at $\lambda=0$
and the rescaled intensity $w$ increases indefinitely when
$\lambda\to 0$. As is shown in Fig.~3, dependence of $T_c$ on 
$q_c$, that is proportional to $h^{2/3}$, reveals more complicated
non-monotonous behavior at $w$ just above its critical value.
The critical temperature becomes an increasing function of the
random field intensity at sufficiently large~$w$.

In solving the equations (\ref{eq:55},\ref{eq:57}) one has to handle
nonuniqueness of the solutions. For example, changing the sign of the
radical in the expression for the critical temperature at $h^2=0$  
(\ref{eq:67}) gives another solution that corresponds to 
the nonphysical branch and has wrong behavior, $T_c\to\infty$, 
in the limit $\lambda\to 0$.
Note that choosing this branch gives another physically absurd
result that $q_h$ grows as temperature increases at $T>T_c$.

Let us summarize the results of the paper.
It is shown that the dynamical action of thermodynamic system
with quenched disorder can be calculated as 
a second Legendre transformation of the effective free energy
functional. Despite the technique was employed to study
the disordered system in the mean field approximation, it
can be equally applied for consideration of fluctuational effects
when going beyond the scope of the mean field theory.
From the other hand, owing to the reflection symmetry of the system,
$\phi\to -\phi$, we have eliminated
$\langle\phi\rangle$ from the considerations concerning
the ergodicity breaking transitions in the case of symmetric 
distribution of quenched variables. Clearly, when the
distribution is nonsymmetric or $A_3\ne 0$ (the presence of
cubic anharmonicity), $\langle\phi\rangle\ne 0$ and 
the dynamical action  contains the terms dependent on
the averaged order parameter. Eqs.~(\ref{eq:44}) then
give coupled equations  for the order parameter
and correlation functions.

The method is applied for the study of $\phi^4$ model of
thermodynamic system with quenched couplings and external field
written in the site representation.
Asymptotics of correlator is found to be affected by 
the random static field. 
It is characterized by the field induced parameter
$q_h$ which is a decreasing function of temperature at $T>T_c$.
Discussion at the end of Sec.~4 led us to the conclusion
that the difference between the memory parameter $q$ and $q_h$
plays the role of an order parameter, so that adiabatic and isothermal
susceptibilities differ if $\Delta q=q-q_h\ne 0$. 
Numerical analysis reveals that 
using the system (\ref{eq:54}) below $T_c$ would 
predict discontinuity of the susceptibility at the critical point.
Much more reasonable results can be obtained in the case of weak ergodicity
breaking. So aging  plays an important part in 
dynamics of the system below critical point.
Recently reported results~\cite{Pitard} for heteropolymers
support the conclusion.

\newpage  

\section*{Appendix A}
\setcounter{section}{1}
\setcounter{equation}{0}
\renewcommand{\theequation}{\Alph{section}.\arabic{equation}}
In the bulk of the paper the solutions under investigation
are casual. In general, this is not the case. An instanton
motion, where both initial and final boundary conditions
need to be fixed, provides an important example of this kind. 
In this Appendix we show that
the mapping, recently introduced in~\cite{Lopatin}, 
between the uphill motion
related to an instanton and the corresponding downhill motion
going back in time can be constructed by making use of SUSY formalism.   

Let us define the transformation
of superfields as follows:
\begin{equation}
  \label{eq:A1}
  \phi(z)\rightarrow \tilde{\phi}(z)=T_{-}(z)\phi(z)\, ,\qquad
  \tilde{\phi}(z)\rightarrow \phi(z)=T_{+}(z)\tilde\phi(z)\, ,
\end{equation}
where
\begin{equation}
\label{eq:A2}
T_{\pm}(z)\equiv\exp(\pm\bar{\theta}\theta\,\partial_{t})\, .
\end{equation}
Inserting Eq.~(\ref{eq:A1}) into 
the action (\ref{eq:8},\ref{eq:9}) integrated from 
$t=t_i$ to $t=t_f$ gives the expression in terms of
superfields $\tilde\phi$
\begin{subequations}
\label{eq:A3}
\begin{eqnarray}
S&=&\tilde{S}+(V_f-V_i)/T\, ,\quad
\tilde{L}=\bar{D}_{-}\tilde{\phi}D_{-}\tilde{\phi}+V(\tilde{\phi})\, ,
\label{eq:A3a}\\
\bar{D}_{-}&=&T_{-}(z)\bar{D}T_{+}(z)=
\frac{\partial}{\partial\theta}+\bar{\theta}
\frac{\partial}{\partial t}\equiv D^{\prime}_{t\to -t}\, ,
\label{eq:A3b}\\
D_{-}&=&T_{-}(z)DT_{+}(z)=
\frac{\partial}{\partial\bar{\theta}}
\equiv \bar{D}^{\prime}_{t\to -t}\, ,
\label{eq:A3c}
\end{eqnarray}
\end{subequations}
where $V_i\equiv V(\eta(t_i))$ and $V_f\equiv V(\eta(t_f))$.
From Eq.~(\ref{eq:A1}) it is clear that
\begin{equation}
  \label{eq:A4}
  \vc{G}(z_1,z_2)=T_{+}(z_1)T_{+}(z_2)\tilde\vc{G}(z_1,z_2)\, , \quad
\tilde\vc{G}(z_1,z_2)=\langle\tilde\phi(z_1)\tilde\phi(z_2)\rangle\,.
\end{equation}
The latter follows the relations:
\begin{subequations}
\label{eq:A5}
\begin{eqnarray}
\langle\varphi(t_1)\varphi(t_2)\rangle&=&
\frac{\partial^2}{\partial t_1\partial t_2}\tilde C(t_1,t_2)+
\frac{\partial}{\partial t_1}\tilde G(t_1,t_2)+
\frac{\partial}{\partial t_2}\tilde G(t_2,t_1)\, ,
\label{eq:A5a}\\
G(t_1,t_2)&=&
\tilde G(t_1,t_2)+\frac{\partial}{\partial t_2}\tilde C(t_1,t_2)\, ,
\label{eq:A5b}\\
C(t_1,t_2)&=&\tilde C(t_1,t_2)\, .
\label{eq:A5c}
\end{eqnarray}
\end{subequations}
It remains to notice that FDT is invariant under the action of
the transformation (\ref{eq:A1}) followed by the inversion of time
$t\to -t$, so the Lagrangian $\tilde L$ describes normal downhill
motion formally inverted in time. The resulting relations (\ref{eq:A5})
are identical with those derived in~\cite{Lopatin} and enable calculating of
the Green function for the instanton process provided that
the corresponding causal solutions are known. 
More details on the subject will be published elsewhere.

\newpage
\begin{center}
  {\bfseries FIGURE CAPTIONS}
\end{center}

\begin{description}
\item[Figure 1.]
Dependence of the critical value of quenched disorder intensity
$w_c$ on the random field $q_c=[h^2/2]^{1/3}$.
($w_c$ and $h^2$ are calculated in units of $\lambda^2/(6\mu^2)$
and  $\lambda^2/(6\mu^3)$, respectively.)

\item[Figure 2.]
Dependencies of the critical temperatures $T_c$ 
(ergodicity breaking transition) and $T_f$ (freezing transition)
on the quenched disorder intensity $w$ in the absence of random
field, $h^2=0$.
($w$ and the temperatures  
are calculated in units of $\lambda^2/(6\mu^2)$
and  $\mu^2/(\sqrt{6}\lambda)$, respectively.)

\item[Figure 3.]
Temperature of ergodicity breaking transition $T_c$
versus  the intensity of random field $q_c=[h^2/2]^{1/3}$
at various values of $w$.

\end{description}

\newpage
\includegraphics[bb=88 120 530 770]{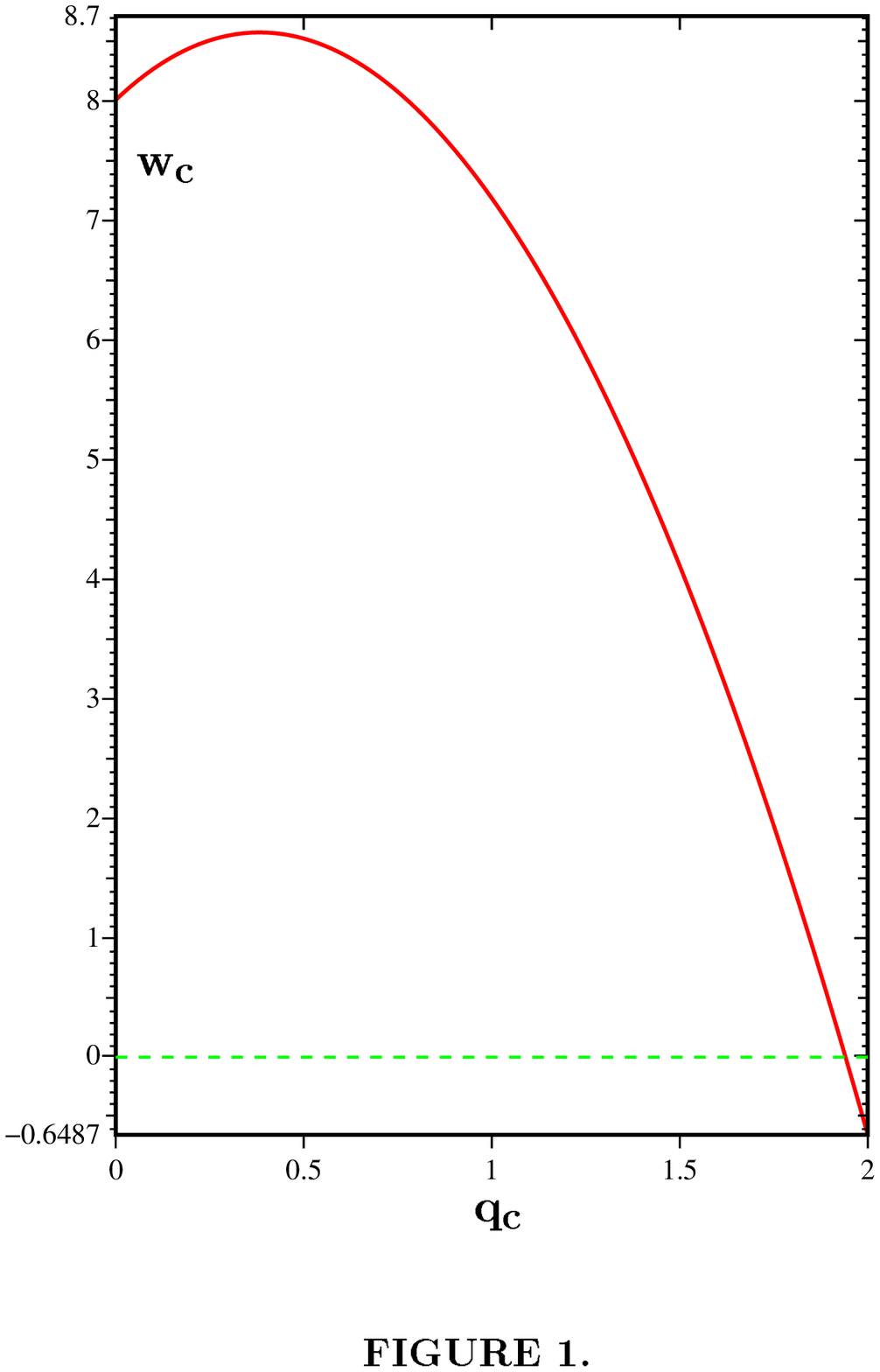}
\newpage
\includegraphics[bb=88 120 530 770]{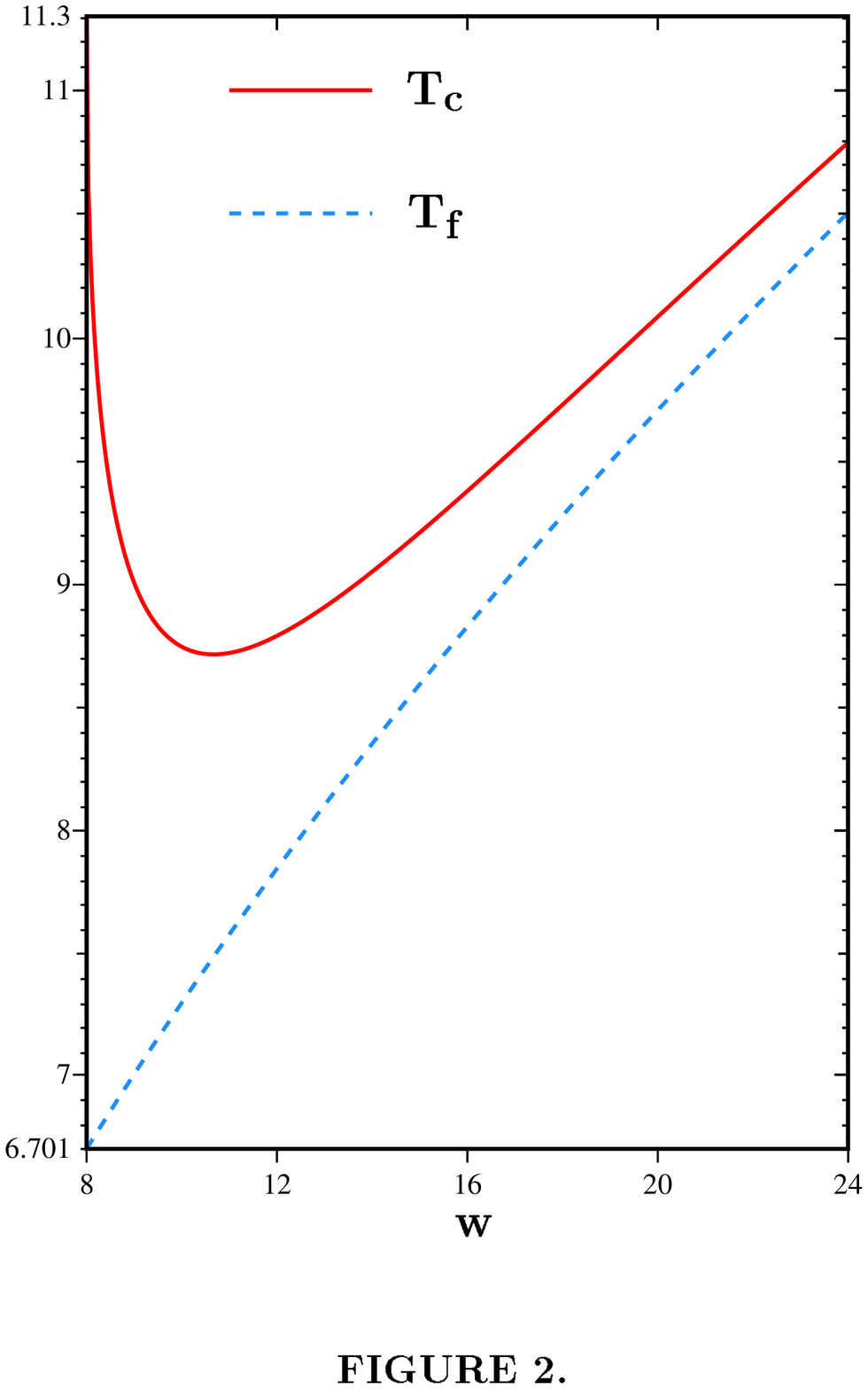}
\newpage
\includegraphics[bb=88 120 530 770]{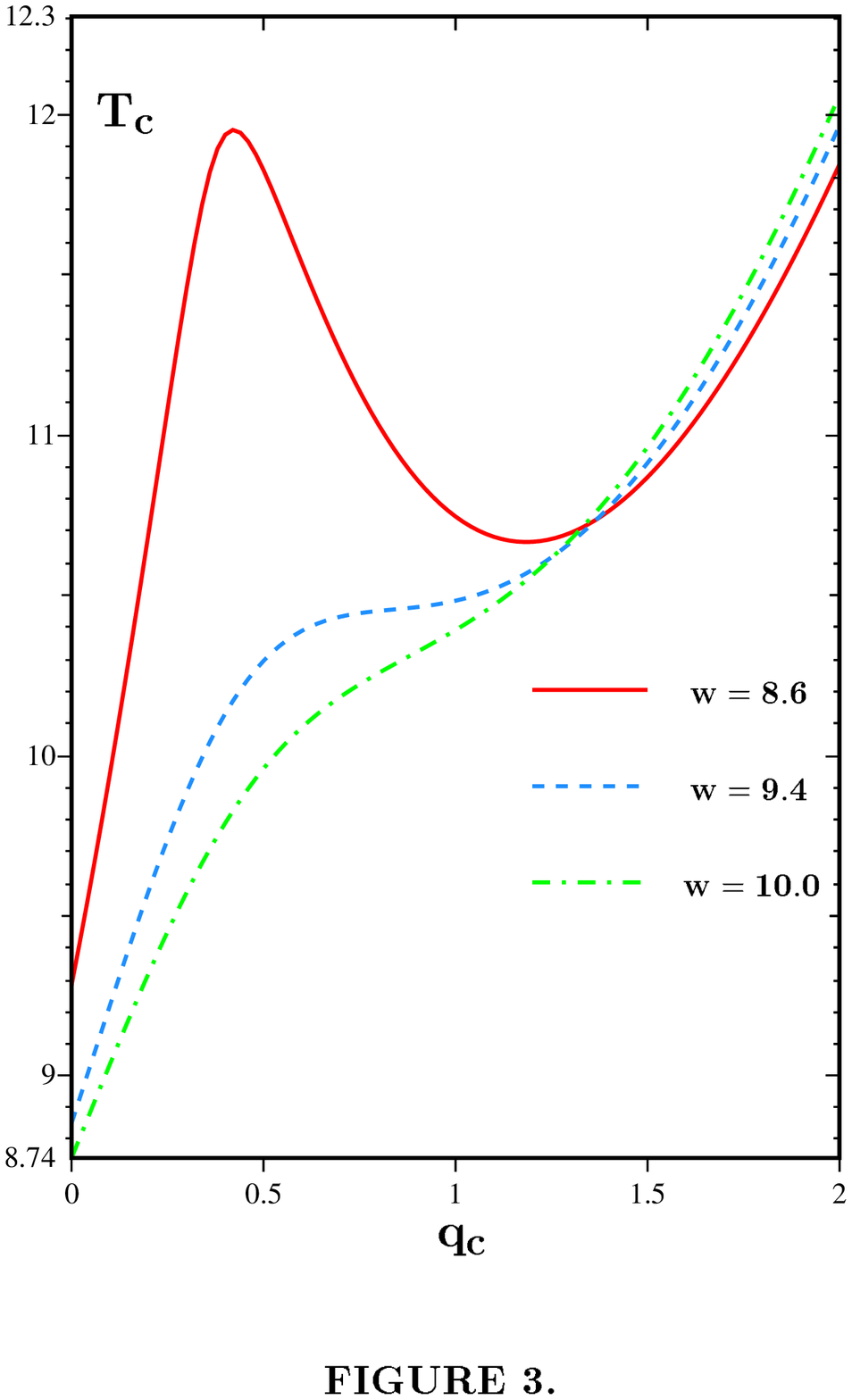}

\end{document}